*Article*

# Experimental Search for Neutron–Antineutron Oscillation with the Use of Ultra-Cold Neutrons Revisited

**Tatsushi Shima**

Research Center for Nuclear Physics, The University of Osaka, Osaka 567-0047, Japan; shima@rcnp.osaka-u.ac.jp

**Abstract**

Neutron–antineutron oscillation (nnbar-osc) is a baryon number-violating process and a sensitive probe for physics beyond the standard model. Ultra-cold neutrons (UCNs) are attractive for nnbar-osc searches because of their long storage time, but earlier analyses indicated that phase shifts on wall reflection differ for neutrons and antineutrons, leading to severe decoherence and a loss of sensitivity. Herein, we revisit this problem by numerically solving the time-dependent Schrödinger equation for the two-component n/nbar wave function, explicitly including wall interactions. We show that decoherence can be strongly suppressed by selecting a wall material whose neutron and antineutron optical potentials are nearly equal. Using coherent scattering length data and estimates for antineutrons, we identify a Ni–Al alloy composition that matches the potentials within a few percent while providing a high absolute value, enabling long UCN storage. With such a bottle and an improved UCN source, the sensitivity could reach an oscillation period $\tau_{nnbar}$ of the order $10^{10}$ s, covering most of the range predicted with certain grand-unified models. This approach revives the feasibility of high-sensitivity nnbar-osc searches using stored UCNs and offers a clear path to probe baryon-number violation far beyond existing limits.

## 1. Introduction

Neutron–antineutron oscillation (hereafter "nnbar-osc") is the process in which a neutron is converted to its antiparticle, meaning the violation of the baryon number by two units, which is not allowed in the standard model of particle physics or in the minimal SU(5) model of the grand unified theory (GUT). SO(10), a GUT model, the next simplest model to SU(5), allows nnbar-osc, but predicts extremely long (>$10^{38}$ y) periods [1]. More frequent nnbar-osc with a shorter (around $10^{10}$ s) oscillation period $\tau_{nnbar}$ is predicted by grand unified theories (GUTs) that contain a left–right symmetric structure in their intermediate gauge groups. Examples include the models of $SU(4)_C \times SU(2)_L \times SU(2)_R$, SO(10), and SUSY models associated with the E6 group, with a symmetry-breaking energy scale of around 100 TeV [2]. These models are of particular interest because they are compatible with explanations of the mass spectrum of light neutrinos using the see-saw mechanism and provide candidates for primordial baryogenesis scenarios [3]. More recent theories beyond the standard model based on extra dimensions also suggest a similar region of $\tau_{nnbar}$[4,5]. Therefore, experimental observation of nnbar-osc provides important information for selecting models beyond the standard model. Since $\tau_{nnbar}$ predicted by those theories ranges from ~$10^8$ s to ~$10^{10}$ s [2–5], experimental exploration of nnbar-osc with $\tau_{nnbar}$ up to ~$10^{10}$ s is highly demanded. The current best limit on $\tau_{nnbar}$ for free neutrons is $\tau_{nnbar}$ > 8.6 × $10^7$ s (90% C.L.), obtained from the detection of antineutrons produced in flight after a flight time of approximately 0.1 s by cold neutrons provided at the ILL research reactor [6]. A new experiment based on the same principle but offering a thousand-fold improvement in discovery potential is planned at the European Spallation Source (ESS) [7]. An experimental search for the exotic decay of $^{16}$O nuclei, arising from the conversion of an intra-nuclear neutron to an antineutron, has been performed with Super Kamiokande (SK); this sets the limit of $\tau_{nnbar}$ > 4.7 × $10^8$ s (90% C.L.) after the correction for the suppression effect of the nuclear field [8]. Since the next-generation Hyper–Kamiokande (HK) detector will have a fiducial volume five times larger than SK, a sensitivity of 1.4 × $10^9$ s is expected. The sensitivities of those planned experiments are, however, not sufficient to fully cover the theoretically preferred region of $\tau_{nnbar}$, and therefore a new experimental plan with higher sensitivity is needed. The sensitivity of the experiment to nnbar-osc is determined by the conversion probability $P_{nnbar}$ from neutron to antineutron. Here, $P_{nnbar}$ is obtained as the solution of Equation (1), which is the Bloch equation for the two-component wave function for neutron-antineutron systems under a sufficiently reduced external magnetic field.

$$i\hbar\frac{\partial}{\partial t}\begin{pmatrix}\psi_n(t)\\ \psi_{nbar}(t)\end{pmatrix}=\begin{pmatrix}E_n-i\Gamma_\beta/2 & \varepsilon\\ \varepsilon & E_{nbar}-i\Gamma_\beta/2\end{pmatrix}\begin{pmatrix}\psi_n(t)\\ \psi_{nbar}(t)\end{pmatrix}\quad, \tag{1}$$

where $E_n$ and $E_{nbar}$ are the kinetic energies of the neutron and antineutron and should be equal to each other, if the CPT symmetry is conserved during free flight. $\Gamma_\beta$ stands for the width of the neutron β decay and is equal to 1.138 × $10^{-3}$ [/s], adopting the CODATA value of the neutron lifetime, 878.4 s [9]. $\varepsilon$ denotes the vacuum expectation value of the interaction causing neutron-antineutron conversion, and its inverse is related to the oscillation period $\tau_{nnbar}$ as $\varepsilon = \hbar/\tau_{nnbar}$. The probability of finding an antineutron at the time $t$ starting with the initial condition of pure neutron, i.e., $\psi_n(t=0)=1$ and $\psi_{nbar}(t=0)=0$, is then obtained as

$$P_{nnbar}=\left|\psi_{nbar}(t)\right|^2\cong(\varepsilon/\hbar)^2 t^2\quad, \tag{2}$$

if $t$ is much smaller than $\tau_{nnbar}$. The concept of using UCNs for nnbar-osc searches was proposed in the early 1980s [10–13], motivated by the possibility of extending the observation time by several orders of magnitude compared with beam experiments. Subsequent detailed analyses, however, indicated that differences in the wall-interaction potentials for neutrons and antineutrons could lead to phase shifts upon reflection, thereby suppressing the oscillation probability. Under the assumptions made at the time, this suppression was estimated to be so large that the expected sensitivity gain would be minimal, and the approach was not pursued further. Those early estimates essentially compared only the energy scales of the wall potentials and the oscillation term, without explicitly taking into account the short dwelling time (~10 ns [14,15]) of the UCNs in the wall. Recognizing that the dwelling time can significantly influence the phase evolution motivated the present re-evaluation.

$$i\hbar\frac{\partial}{\partial t}\begin{pmatrix}\psi_n(t)\\ \psi_{nbar}(t)\end{pmatrix}=\begin{pmatrix}E_n-i\Gamma_\beta/2+U_n & \varepsilon\\ \varepsilon & E_{nbar}-i\Gamma_\beta/2+U_{nbar}\end{pmatrix}\begin{pmatrix}\psi_n(t)\\ \psi_{nbar}(t)\end{pmatrix}\quad, \tag{3}$$

where $U_n$ ($U_{nbar}$) is the potential of the interaction of a neutron (an antineutron) with matter, which is obtained as the Fermi's pseudopotential with the parameter of the coherent nuclear scattering of the nuclei contained in the material of the wall. The solution of Equation (3) is obtained as Equation (4) [13]

$$
\begin{aligned}
P_{nnbar} &= \left|\psi_{nbar}(t)\right|^2 = \frac{4(\varepsilon/\hbar)^2}{\omega_W^2 + 4(\varepsilon/\hbar)^2} \cdot \exp\left(-\left(\Gamma_\beta + \Gamma_s\right)\frac{t}{\hbar}\right) \cdot \sin^2\left(\frac{1}{2}\sqrt{\omega_W^2 + 4(\varepsilon/\hbar)^2}\,t\right) \\
&= \frac{4(\varepsilon/\hbar)^2}{\nu^2} \cdot \exp\left(-\left(\Gamma_\beta + \Gamma_s\right)\frac{t}{\hbar}\right) \cdot \sin^2\left(\frac{\nu}{2}t\right) \\
\varepsilon &\equiv \frac{\hbar}{\tau_{nnbar}}\ ,\ \omega_W \equiv \frac{U_n - U_{nbar}}{\hbar}\ ,\ \Gamma_\beta \equiv \frac{\hbar}{\tau_\beta}\ ,\ \Gamma_s \equiv \frac{\hbar}{\tau_s}\ ,
\end{aligned}
\tag{4}
$$

where $\omega_W$ and $\nu$ are the angular frequencies defined in Equation (4). $\tau_s$ is the neutron storage time of the bottle. In a vacuum, $\omega_W$ is 0, $t$ is considered to be approximately 0.5 s, which is the mean free time of neutrons $t_{free}$, and $\varepsilon/\hbar$ is smaller than ~$5 \times 10^{-9}$ [rad·Hz]. Therefore, Equation (4) is well approximated by Equation (2). In the medium, since the strength of $U_n$ or $U_{nbar}$ is typically of the order of $10^{-7}$ [eV], if no adjustment is made to match $U_n$ and $U_{nbar}$, $\omega_W$ is expected to be approximately $1.52 \times 10^8$ [rad·Hz], being more than 16 orders of magnitude larger than $\varepsilon/\hbar$, resulting in $P_{nnbar}$ being smaller than ~$10^{-30}$ if $\sin(\nu t/2)$ is not small, as discussed in [15]. In this case, the n/nbar wave function should be reset in reflections on the wall. Therefore, if we perform a measurement with a storage time of 1000 s and an average free flight time of UCNs in the bottle of 0.5 s, it is equivalent to repeating a measurement with $t = 0.5$ s 2000 times. This improves the sensitivity of the ILL experiment not by a factor of 2000 but only by the square root of 2000, leading to a sensitivity of $\tau_{nnbar}$~$10^9$ s. This sensitivity is still lower than the expected sensitivities of future HK or ESS experiments. This situation is enough to discourage people. After [10–13], this problem was not considered further. However, it should be noted that the time $t$ appearing in Equation (4) is the dwelling time of a neutron or an antineutron in the wall $t_W$ when they are traveling in matter, and $t_W$ is known to be ~10 ns [14,15], much shorter than $t_{free}$ or $\tau_s$. $t_W$ provides $\nu t$~1.52 rad, meaning the factor $\sin(\nu t/2)$ is not necessarily large, depending on the combination of $U_n$ and $U_{nbar}$, and a more detailed treatment is required to evaluate $P_{nnbar}$. In this work, the influence of the reflection on the wall is taken into account explicitly to provide a realistic evaluation of $P_{nnbar}$. This work points out that, by selecting or engineering a wall material whose coherent scattering lengths for neutrons and antineutrons are nearly equal, $\Delta U \equiv U_n - U_{nbar}$ can be reduced by orders of magnitude, suppressing $\omega_W t_W$ and recovering the oscillation sensitivity. In what follows, the method of numerical calculation and the result are presented.

## 2. Method for Realistic Evaluation of $P_{nnbar}$

The precise time evolution of the n/nbar wave function can be computed by explicitly multiplying the Hamiltonian in Equation (3) by the initial wave function. The wave functions before and after the reflection on the wall are connected as

$$
\begin{aligned}
\begin{pmatrix} \psi_n(t_W + t) \\ \psi_{\bar{n}}(t_W + t) \end{pmatrix} &= \exp\left[-\left(iE_n + \frac{\Gamma_\beta}{2}\right)\frac{t_W}{\hbar}\right] \cdot \begin{pmatrix} \cos\nu t_W + \dfrac{i\omega_W}{2\nu}\sin\nu t_W & -\dfrac{i\varepsilon}{\hbar\nu}\sin\nu t_W \\ -\dfrac{i\varepsilon}{\hbar\nu}\sin\nu t_W & \cos\nu t_W - \dfrac{i\omega_W}{2\nu}\sin\nu t_W \end{pmatrix} \begin{pmatrix} \psi_n(t) \\ \psi_{\bar{n}}(t) \end{pmatrix} \\
&= \exp\left[-\left(iE_n + \frac{\Gamma_\beta}{2}\right)\frac{t_W}{\hbar}\right] \cdot \mathbf{R} \begin{pmatrix} \psi_n(t) \\ \psi_{\bar{n}}(t) \end{pmatrix} .
\end{aligned}
\tag{5}
$$

Here, **R** is the reflection matrix defined as

$$\mathbf{R} \equiv \begin{pmatrix} \cos \nu t_W + \frac{i\omega_W}{2\nu} \sin \nu t_W & -\frac{i\varepsilon}{\hbar \nu} \sin \nu t_W \\ -\frac{i\varepsilon}{\hbar \nu} \sin \nu t_W & \cos \nu t_W - \frac{i\omega_W}{2\nu} \sin \nu t_W \end{pmatrix} . \tag{6}$$

The effect of the multiple reflections can be evaluated by multiplying **R** by the number of reflections. Meanwhile, the time evolution of the wave function during the travel from wall to wall is taken into account by introducing another matrix **F**, provided by setting $U_n = U_{nbar} = 0$, $\omega_W = 0$ in Equation (6):

$$\mathbf{F} \equiv \begin{pmatrix} \cos\left(\varepsilon t_{free}/\hbar\right) & -i\sin\left(\varepsilon t_{free}/\hbar\right) \\ -i\sin\left(\varepsilon t_{free}/\hbar\right) & \cos\left(\varepsilon t_{free}/\hbar\right) \end{pmatrix} . \tag{7}$$

Here, $t_{free}$ is the mean free time between reflections on the walls, obtained by $t_{free} = t/m$ using the total elapsed time $t$ and the total number of reflections $m$. Using matrices **F** and **R**, the wave function after $m$ number of reflections is given as

$$\begin{aligned} \begin{pmatrix} \psi_n(t) \\ \psi_{nbar}(t) \end{pmatrix} &= \exp\left[-\left(iE_n + \frac{\Gamma_\beta}{2}\right)\frac{t}{\hbar}\right] \cdot (\mathbf{FR})^m \cdot \begin{pmatrix} \psi_n(t=0) \\ \psi_{nbar}(t=0) \end{pmatrix} \\ &= \exp\left[-\left(iE_n + \frac{\Gamma_\beta}{2}\right)\frac{t}{\hbar}\right] \cdot \frac{D^2 - b^2}{2Da\sin\left(\varepsilon t_{free}/\hbar\right)} \begin{pmatrix} 1 & 1 \\ \frac{-a\sin\left(\varepsilon t_{free}/\hbar\right)}{D-b} & \frac{a\sin\left(\varepsilon t_{free}/\hbar\right)}{D+b} \end{pmatrix} \\ &\times \begin{pmatrix} (a+iD)^m & 0 \\ 0 & (a-iD)^m \end{pmatrix} \begin{pmatrix} \frac{a\sin\left(\varepsilon t_{free}/\hbar\right)}{D+b} & -1 \\ \frac{a\sin\left(\varepsilon t_{free}/\hbar\right)}{D-b} & 1 \end{pmatrix} \begin{pmatrix} \psi_n(0) \\ \psi_{nbar}(0) \end{pmatrix} . \end{aligned} \tag{8}$$

Here, it should be noted that the number of reflections and the number of free flights from wall to wall are the same, $m$. In Equation (8), the new parameters $a$, $b$, and $D$ are introduced for easy understanding by using the following definitions:

$$a \equiv \cos \nu t_W \ , \ b \equiv \sin \nu t_W \ , \ D \equiv \sqrt{a^2 \sin^2\left(\varepsilon t_{free}/\hbar\right) + b^2} \cong b$$

Starting with the initial condition of a pure neutron,

$$\begin{pmatrix} \psi_n(0) \\ \psi_{nbar}(0) \end{pmatrix} = \begin{pmatrix} 1 \\ 0 \end{pmatrix} ,$$

the probability of finding an antineutron $P_{nnbar}$ at time $t$ is obtained as

$$\begin{aligned} P_{nnbar}(t) = \left|\psi_{nbar}(t)\right|^2 &= \exp\left(-\left(\Gamma_\beta + \Gamma_s\right)t\right)\left(\frac{a\sin\left(\varepsilon t_{free}/\hbar\right)}{2D}\right)^2 \\ &\times\left[-(a-iD)^m + (a+iD)^m\right]\left[-(a+iD)^m + (a-iD)^m\right] . \end{aligned} \tag{9}$$

Here,

$$
\begin{aligned}
&(a \pm iD)^m \cong a^m \left(1 \pm im\frac{b}{a}\right),\\
&P_{nnbar}(t) \cong \exp\left(-\left(\Gamma_\beta + \Gamma_s\right)t\right)\left(\frac{a\sin\left(\varepsilon t_{free}/\hbar\right)}{2D}\right)^2 a^{2m}\\
&\times\left[\left(1+im\frac{b}{a}\right)-\left(1-im\frac{b}{a}\right)\right]\left[\left(1-im\frac{b}{a}\right)-\left(1+im\frac{b}{a}\right)\right]\\
&\cong 4\exp\left(-\left(\Gamma_\beta + \Gamma_s\right)t\right)\left(\frac{a\sin\left(\varepsilon t_{free}/\hbar\right)}{2b}\right)^2 a^{2m} m^2 \left(\frac{b}{a}\right)^2\\
&=\exp\left(-\left(\Gamma_\beta + \Gamma_s\right)t\right)\cdot a^{2m}\left(m\sin\left(\varepsilon t_{free}/\hbar\right)\right)^2 \cong \exp\left(-\left(\Gamma_\beta + \Gamma_s\right)t\right)\cdot a^{2m}\left(\varepsilon m t_{free}/\hbar\right)^2\\
&=\exp\left(-\left(\Gamma_\beta + \Gamma_s\right)t\right)\cdot a^{2m}\left(\varepsilon t/\hbar\right)^2 = \exp\left(-\left(\Gamma_\beta + \Gamma_s\right)t\right)\cdot a^{2m}\left(t/\tau_{nnbar}\right)^2\\
&=\exp\left(-\left(\Gamma_\beta + \Gamma_s\right)t\right)\cdot S_{amp}^2\left(t/\tau_{nnbar}\right)^2 = \exp\left(-\left(\Gamma_\beta + \Gamma_s\right)t\right)\cdot S_{prob}\left(t/\tau_{nnbar}\right)^2,
\end{aligned}
\tag{10}
$$

where $S_{amp} \equiv a^m$ is the amplitude-level suppression factor, and $S_{prob} \equiv S_{amp}^2 = a^{2m}$ is the probability-level suppression factor. If one succeeds in adjusting $U_n$ and $U_{nbar}$ within 3% difference, $\nu t_W \sim 0.045$ and $a = 0.999$, $S_{amp}$ is 0.60 in case of 500 reflections. This amplitude-level suppression factor is within the feasibility of the nnbar-osc experiment using the UCN bottle. For example, if a UCN enters the bottle with the dimension of 5 m (width) × 5 m (depth) × 1 m (height) at the height of 0.5 m with the initial velocity of 5.0 m/s perpendicular to the side wall, as shown schematically in Figure 1, the UCN bounces off the floor every 0.64 s and is reflected off the side wall every 1.0 s. In this case, the number of reflections $m$ is given as $m = t/0.64 + t/1.0 = 2.563t$, and the probabilities $P_{nnbar}(t)$ for the examples of a 1%, 3%, and 100% difference between $U_n$ and $U_{nbar}$ and $\Gamma_s = 500$ s and $\tau_{nnbar} = 10^9$ s are obtained as shown in Figure 2.

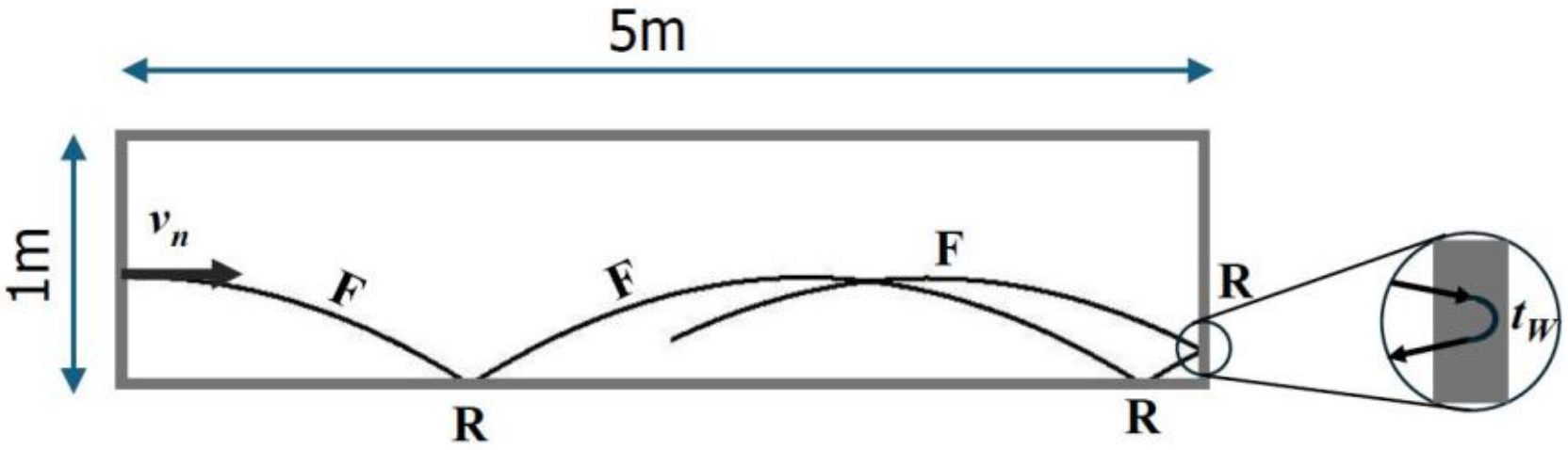

**Figure 1.** Schematic side view of the experimental setup for the nnbar-osc using UCN considered in the present work. A UCN enters from the left side of the bottle at a height of 0.5 m with a velocity of 5.0 m/s perpendicular to the wall. Parabolic curves show the trajectories of the UCN. The time evolution of the wave function of the UCN is governed by the operators **F** and **R**. In the real experiment, the bottle will be surrounded by the detectors for multiple pions emitted from the annihilation of antineutrons in the walls.

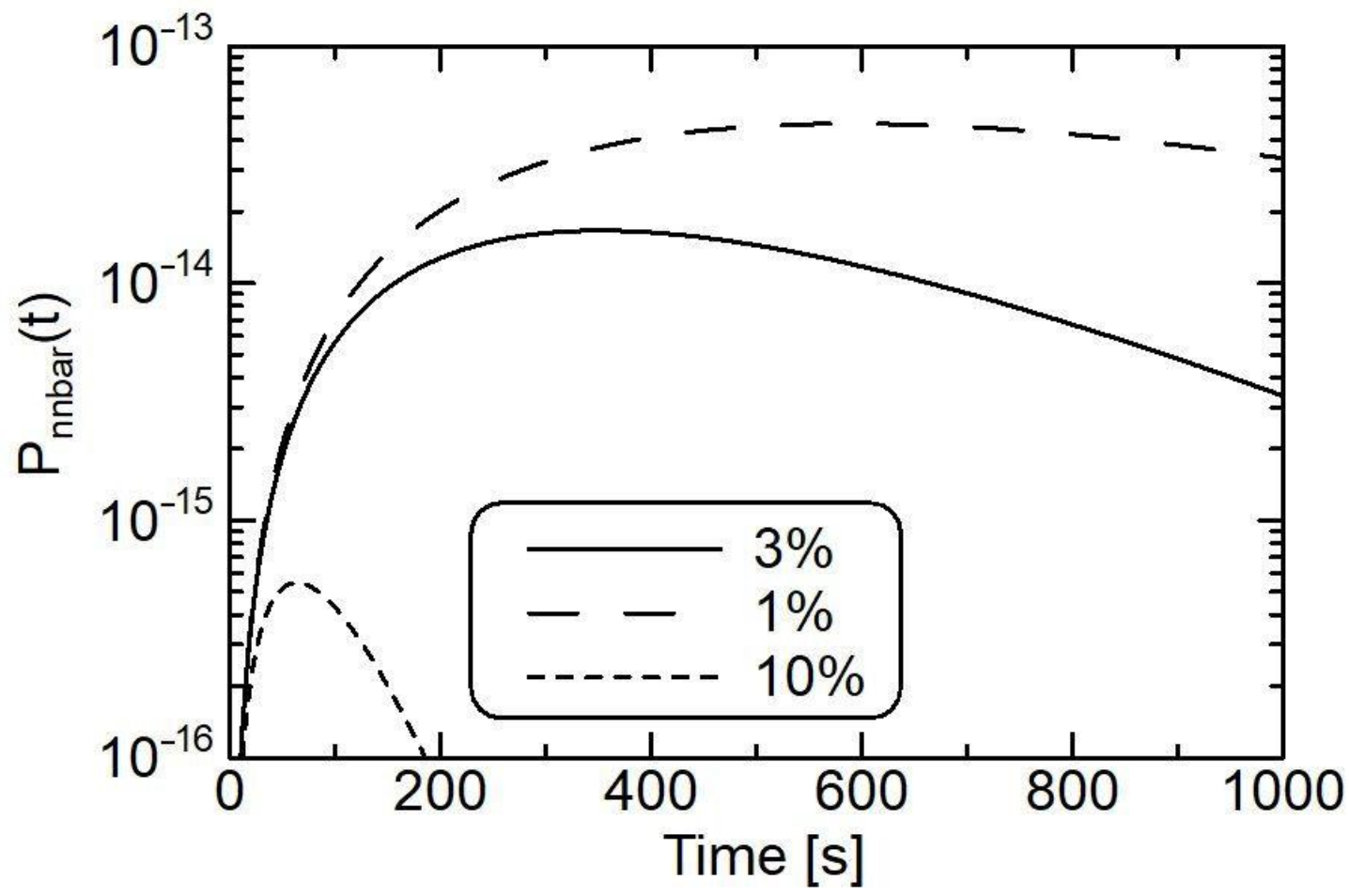

**Figure 2.** Time dependences of the probability $P_{nnbar}$ for different potential differences between $U_n$ and $U_{nbar}$; solid, long-dashed, and short-dashed curves indicate the cases of relative differences of 3%, 1%, and 10%, respectively. The storage time of the bottle $\tau_s$ was assumed to be 500 s.

As shown in Figure 2, the time dependence of $P_{nnbar}$ is characterized by an increase due to the $t^2$ term and a decrease due to $S_{prob}$, the neutron β-decay, and the storage time of the bottle.

## 3. Results

As discussed above, it is essential to adjust the neutron-wall potential $U_n$ and the antineutron-wall potential $U_{nbar}$ within an accuracy of a few percent to obtain a high sensitivity to nnbar-osc with the use of a UCN bottle. In addition, to increase the storage efficiency of the bottle, it is important to increase the absolute values of $U_n$ and $U_{nbar}$ as much as possible. The potentials are given as Fermi's pseudopotential in terms of the real parts of the coherent scattering length $b_{coh_n}$ and $b_{coh_nbar}$ as

$$U_n = \frac{2\pi\hbar^2}{m_n}\rho_A \operatorname{Re}\left(b_{coh_n}\right)$$
$$U_{nbar} = \frac{2\pi\hbar^2}{m_n}\rho_A \operatorname{Re}\left(b_{coh_nbar}\right) \quad . \tag{11}$$

Here, $b_{coh_n}$ and $b_{coh_nbar}$ stand for neutron-nuclear coherent scattering and anti-neutron-nuclear coherent scattering, respectively. $\rho_A$ denotes the number density of nuclides with the mass number $A$; $m_n$ is the neutron rest mass. At present, almost no data are available for $b_{coh_nbar}$ compared to $b_{coh_n}$. However, based on the analysis of existing antiproton-nucleus scattering/absorption data, as well as anti-hydrogen, the coherent antineutron scattering length $b_{coh_nbar}$ was evaluated as

$$b_{coh_nbar}\left(A\right) = \left(1.54 \pm 0.03\right) \cdot A^{0.311 \pm 0.05} - \left(1.00 \pm 0.04\right) i \ [\text{fm}] \tag{12}$$

as a function of the nuclear mass number $A$ with a precision of 3% [16]. Compared to the existing data of $b_{coh_n}$, $^{96}$Mo is a candidate isotope suitable for the material of the bottle. $b_{coh_n}$ and $b_{coh_nbar}$ of $^{96}$Mo are 6.20 fm and 6.37 fm, respectively, and the corresponding pseudopotentials are $U_n$ = 146.3 neV and $U_{nbar}$ = 150.3 neV, whose relative difference is 2.7%. As mentioned above, since a higher $U_n$ and $U_{nbar}$ is more useful, it is necessary to enhance the flexibility in finding a candidate material. This can be achieved by considering compounds or alloys as the bottle material. If one considers a material made of two elements A and B with the stoichiometry of $(1 - x){:}\,x$, the difference between $U_n$ and $U_{nbar}$ is obtained as

$$U_n - U_{nbar} = \frac{2\pi}{m_n}\rho\left[x\left(b_{nA} - b_{nbarA}\right) + (1-x)\left(b_{nB} - b_{nbarB}\right)\right] \quad , \tag{13}$$

which becomes very close to zero when $x$ is chosen as

$$x = \frac{\left(b_{nB} - b_{nbarB}\right)}{\left(b_{nB} - b_{nbarB}\right) - \left(b_{nA} - b_{nbarA}\right)} \quad . \tag{14}$$

Here, it is assumed that the number densities of A and B should be of the same value as the average number density of the alloy or the compound in the case of a homogeneous, uniform, and perfect mixture. The design of such material is represented in Table 1.

**Table 1.** Parameters of the candidate UCN bottle material.

| Elemental Composition | Re ($b_{coh_n}$) [fm] | Re ($b_{coh_nbar}$) [fm] | Molar ratio [%] |
|---|---|---|---|
| Ni | 10.3 | 5.47 | 85.2 |
| Al | 3.449 | 4.29 | 14.8 |

The alloy made of 85.2% Ni atoms and 14.8% Al atoms provides $U_n \sim U_{nbar} \sim$220 neV, which is comparable to the $U_n$ of nickel (243 neV). In addition, Ni-Al alloy does not need expensive enriched isotopes. Therefore, it is considered the best material for the UCN bottle. A similar consideration was made in [17], which proposed a new nnbar-osc experiment using very-cold neutrons (VCNs) transported for a very long distance using a VCN guide made of the isotopic composition of $^{184}$W(87.7%) and $^{186}$W(12.3%), giving $U_n \sim U_{nbar} \sim$106 neV. The presently recommended Ni-Al alloy provides twice the performance of UCN storage with non-expensive raw materials, which is an advantage in constructing large-volume storage. For example, to coat the inner walls of the UCN bottle of the size shown in Figure 1 with tungsten with a thickness of 1 μm, approximately 1.35 kg of enriched tungsten is required, which will cost roughly USD 1~2 M.

## 4. Discussion

Using a UCN bottle made of the material discussed in the previous section, the decoherence effect due to the interaction with the wall can be efficiently suppressed. For example, using the UCN bottle shown in Figure 1, made of a Ni-Al alloy with the parameters listed in Table 1, the averaged free-flight time is approximately 0.39 s for UCNs with a velocity of 5 m/s, assuming a storage time of 500 s; 500 reflections are made on the side wall and 781 reflections on the floor, for a total of 1281 reflections. The loss due to decoherence at the walls is estimated using Equation (11).

Since antineutrons are detected via the multiple pions emitted during their annihilation at the walls, they can be observed at each reflection. The expected number of signals per one UCN is then calculated as follows:

$$\begin{aligned} Y_{nbar} &= \sum_{m=1}^{\infty} P_{nnbar}\left(t_m\right) = \sum_{m=1}^{\infty} \exp\left(-\left(\Gamma_\beta + \Gamma_s\right)t_m\right) \cdot a^{2m}\left(t_m / \tau_{nnbar}\right)^2 \\ &= \sum_{m=1}^{\infty} \exp\left(-\left(\Gamma_\beta + \Gamma_s\right)\frac{m}{2.563}\right) \cdot a^{2m}\left(m / \left(2.563 \cdot \tau_{nnbar}\right)\right)^2 \quad . \end{aligned} \tag{15}$$

In Equation (16), the value 2.563 is the average number of reflections per second under the present conditions, and assuming $a = 0.999$, $\Gamma_\beta = 1/\tau_\beta = 1.138 \times 10^{-3}\ [s^{-1}]$, $\Gamma_s = 1/\tau_s = 0.002\ [s^{-1}]$, where $\tau_s$ is the storage time of the bottle, $Y_{nnbar}$ is numerically calculated as shown in Figure 3 as a function of $\tau_{nnbar}$.

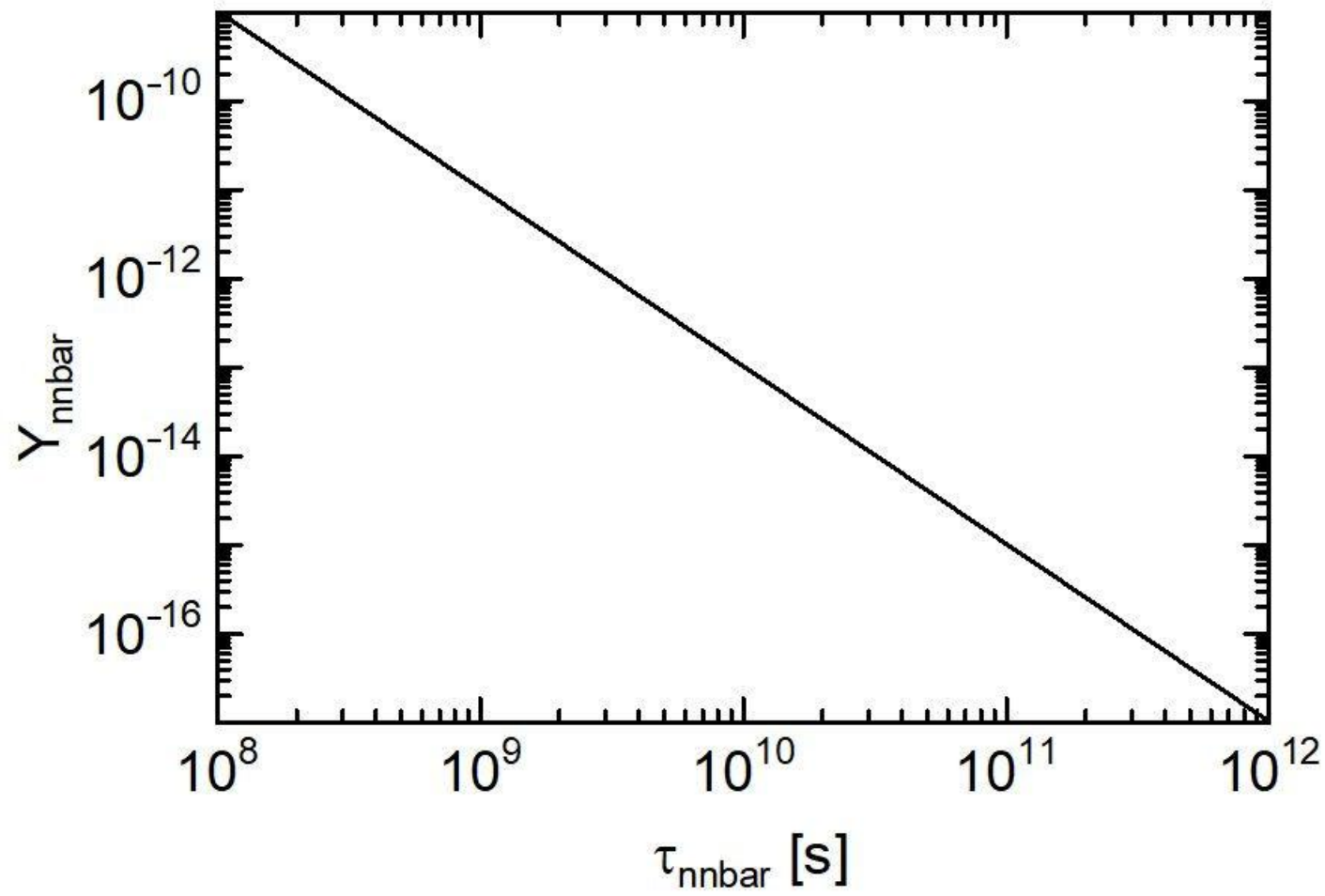

**Figure 3.** Expectation value $Y_{nnbar}$ of the antineutron signal per UCN as a function of the nnbar-osc period $\tau_{nnbar}$. Suppression factor $\alpha$ per reflection was assumed to be 0.999.

The total expected number of antineutron events $Y_{total}$ during the experimental duration $t_{meas}$ is obtained by

$$Y_{total} = \varepsilon \cdot \Phi_{UCN} \cdot Y_{nnbar} \cdot t_{meas} \quad , \tag{16}$$

in terms of the detector efficiency $\varepsilon$ and incident UCN intensity $\Phi_{UCN}$. Assuming the background level is as good as in the case of the ILL experiment, i.e., no event is detected during a one-year measurement, the upper limit on $Y_{total}$ with 90% C.L. is 2.3 events. Assuming the intensity of the UCN beam $\Phi_{UCN}$ is $1.4 \times 10^7$ [/s], which is expected to be achieved by the TUCAN experiment [18], the detection efficiency for antineutron as $\varepsilon$ = 0.5 and $Y_{nnbar}$ up to $7.14 \times 10^{-13}$ can be explored, which corresponds to the 90% C.L. lower limit of $\tau_{nnbar}$ = $3.75 \times 10^9$ s according to Figure 3, which is better than the expected sensitivities in the experiments at HK and ESS.

## 5. Conclusions

In the present work, it was found that a very high sensitivity for nnbar-osc can be achieved using the UCN bottle method by matching the neutron-wall potential and the antineutron-wall potential based on the selection of a suitable material for the wall of the bottle, such as a Ni-Al alloy. With the UCN source for the TUCAN experiment, it is expected that nnbar-osc with an oscillation period up to $\tau_{nnbar}$ ~$3.75 \times 10^9$ s can be explored. A more intense UCN source is proposed for the future research reactor, which is planned to be in operation around 2035 at the site of the closed Monju fast breeder reactor. Since the thermal power of the future research reactor is designed to be 10 MW, being 500 times greater than the beam power of the driver cyclotron of the TUCAN UCN source, an improvement in $\Phi_{UCN}$ by a factor of 10 will be feasible if one employs a refrigerator with sufficient cooling power. This means that the UCN source at the new research reactor can produce $10^8$ UCNs every second. Using such an intense UCN source, nnbar-osc will be explored up to $\tau_{nnbar}$ = $6.66 \times 10^{10}$ s (90% C.L.). This covers almost the entire region of the probability distribution of $\tau_{nnbar}$ predicted by a model beyond the standard model [3]. It should be noted that the present estimation was performed with a fixed velocity (5 m/s, initially horizontal) of UCNs. For a more realistic estimation, it is necessary to conduct a computer simulation with the information on the energy distribution of the stored UCNs. Additionally, the dwelling time $t_W$ of UCN in the wall was fixed to the typical value of 10 ns in the present study. The exact value of $t_W$ can be obtained by taking into account the kinematics of UCNs in the wall; however, the current conclusion regarding the importance of adjusting the neutron-wall and antineutron-wall interaction potential remains unchanged. In fact, independently, the present problem was recently reconsidered by using the exact solution of the full Schrödinger equation and results consistent with those reported here were obtained [19].

**Funding:** This research received no external funding.

**Acknowledgments:** The author thanks H.M. Shimizu from Nagoya University and M. Kitaguchi from KMI, Nagoya University; K. Mishima from RCNP, the University of Osaka; H. Fujioka from the Institute of Science Tokyo; T. Higuchi from the Institute for Integrated Radiation and Nuclear Science, Kyoto University; and S. Kawasaki from KEK/IPNS for fruitful discussions on the design of the experiment using a UCN bottle made of the material with the adjusted potential of the neutron/antineutron interaction with the wall.

**Data Availability Statement**: The original contributions presented in this study are included in the article. Further inquiries can be directed to the corresponding author.

**Conflicts of Interest:** The author declares no conflicts of interest.

## Abbreviations

The following abbreviations are used in this manuscript:

| | |
|---|---|
| UCN | Ultra-cold neutron |
| VCN | Very-cold neutron |
| GUT | Grand unified theory |
| SUSY | Super symmetry |
| nnbar-osc | Neutron-antineutron oscillation |